\documentclass[a4,aps,amsmath,floatfix,nofootinbib,longbibliography,twocolumn,prb,superscriptaddress]{revtex4-1} 
\usepackage{graphicx} 
\usepackage{color}
\usepackage{enumerate} 
\usepackage{braket}
\usepackage[caption=false]{subfig}
\usepackage[colorlinks=true,linkcolor=red]{hyperref}
\usepackage{xcolor}
\newcommand{\bdm}{\begin{displaymath}} 
\newcommand{\edm}{\end{displaymath}} 
\newcommand{\tG}{\tilde{G}}

\newcommand{\pn}{\text{PN}}

\newcommand{\pcsadd}{Center for Theoretical Physics of Complex Systems, Institute for Basic Science (IBS), Daejeon 34126, Republic of Korea}
\newcommand{\iasadd}{New Zealand Institute for Advanced Study, Centre for Theoretical Chemistry and Physics, Massey University, Auckland 0745, New Zealand}

\begin{document}

\title{Taming two interacting particles with disorder}

\author{Diana Thongjaomayum}
\affiliation{\pcsadd}

\author{Alexei Andreanov}
\affiliation{\pcsadd}

\author{Thomas Engl}
\affiliation{\pcsadd}
\affiliation{\iasadd}

\author{Sergej Flach}
\affiliation{\pcsadd}
\affiliation{\iasadd}

\date{\today}

\begin{abstract}
    We compute the scaling properties of the localization length $\xi_2$ of two interacting particles in a one-dimensional chain with diagonal disorder, and the connectivity properties of the Fock states. We analyze record large system sizes (up to $N=20000$) and disorder strengths (down to $W=0.5$). We vary the energy $E$ and the on-site interaction strength $u$. At a given disorder strength the largest enhancement of $\xi_2$ occurs for $u$ of the order of the single particle band width, and for two-particle states with energies at the center of the spectrum, $E=0$. We observe a crossover in the  scaling of $\xi_2$ with the single particle localization length $\xi_1$ into the asymptotic regime for $\xi_1 > 100$ ($W < 1.0$). This happens due to the recovery of translational invariance and momentum conservation rules in the matrix elements of interconnected Fock eigenstates for $u=0$. The entrance into the asymptotic scaling is manifested through a nonlinear scaling function $\xi_2/\xi_1=F(u\xi_1)$.
\end{abstract}


\maketitle

\section{Introduction}

The fundamental question about the interplay of Anderson localization~\cite{anderson1958absence} and many-particle interactions has driven both analytical and numerical studies for decades.~\cite{fleishman1980interaction,shahbazyan1996surface,kozub2000fluctuation,wang2000short,nattermann2003variable, basko2006metal} In particular, celebrated many body localization transitions for a macroscopic system with a finite particle density have been predicted using perturbation theory approaches.~\cite{basko2006metal} It is most natural then to turn attention on a seemingly simple problem of two interacting particles (TIP) in a one-dimensional disordered tight binding chain (with hopping strength $t=1$) which has both ingredients: localization and interaction. While there is little doubt that in one spatial dimension the two particles stay localized for any onsite (and in general short-range) interactions at finite disorder strength, there are conflicting results on how the localization length $\xi_2$ of the most extended TIP states scales with the single particle localization length $\xi_1$ in the limit of weak disorder $\xi_1 \gg 1$. Predictions and numerical conclusions for $\xi_2\propto\xi_1^\alpha$  range from $\alpha=1$~\cite{roemer1997no,roemer1999two} over $\alpha\approx 1.6$ [10] to $\alpha = 2$.~\cite{shepelyansky1994coherent,imry1995coherent} Ponomarev et al suggested that the scaling will be modified with a nonuniversal exponent $\alpha(u)$ depending on the strength of interaction $u$,~\cite{ponomarev1997coherent} and discussed logarithmic corrections $\alpha = a + b \ln (\xi_1)$. Even with only two particles, the computational task turns difficult since weak disorder values are targeted, and the required system size increases rapidly with decreasing disorder strength. The asymptotic scaling sets in for $\xi_1 \geq 100$ when momentum conservation correlations and translational invariance begin to be restored in the single particle eigenstates.~\cite{krimer2015interaction} Yet most numerical studies focused on the more accessible region $ \xi_1 < 100$ and therefore yield at best $\xi_2 \approx 2...3 \xi_1$.~\cite{shepelyansky1994coherent,frahm1995scaling,vonoppen1996interaction,frahm1999interaction,arias1999two,waintal1999two,krimer2011two} A recent Green's function computation by Ref.~\onlinecite{frahm2016eigenfunction} entered the scaling regime with reaching $\xi_1 \approx 400$ and $\xi_2 \approx 9 \xi_1$. However, at the chosen interaction strength unavoidable finite size corrections~\cite{song1997localization} will bring this number down to $\xi_2 \approx 6 \xi_1$ according to our present computations. In addition a completely overlooked impact comes from the interplay of interaction strength $u$ and the eigenstate energy $E$. Also TIP have been studied recently on a three-dimensional lattice in presence of a mobility edge demonstrating a sensitive dependence on the interaction strength.~\cite{stellin2019mobility}

In this work, we show that in the regime of asymptotic scaling $\xi_1 \geq 100$ the connectivity between Fock states (non-interacting eigenstates) is strongly selective due to combining energy conservation with emerging momentum conservation. In order to computationally assess $\xi_2$ in the asymptotic scaling regime, we extend the projected Green's function method used in Ref.~\onlinecite{vonoppen1996interaction} by adding a finite size scaling and significantly increasing the system size compared to the data presented in the literature by more than one order of magnitude up to $N=20000$, and by systematically varying the energy $E$ and the interaction strength $u$. We show that the largest values of $\xi_2$ at given $\xi_1$ are obtained for $E=0$ and $u\approx 3$. We report the record values $\xi_2 \approx 16 \xi_1$. The entrance into the asymptotic scaling is manifested through a non-linear scaling function $\xi_2/\xi_1=F(u\xi_1)$.

\begin{figure*}
    \centering{}\subfloat[]{
    \begin{centering}
        \includegraphics[width=0.33\textwidth]{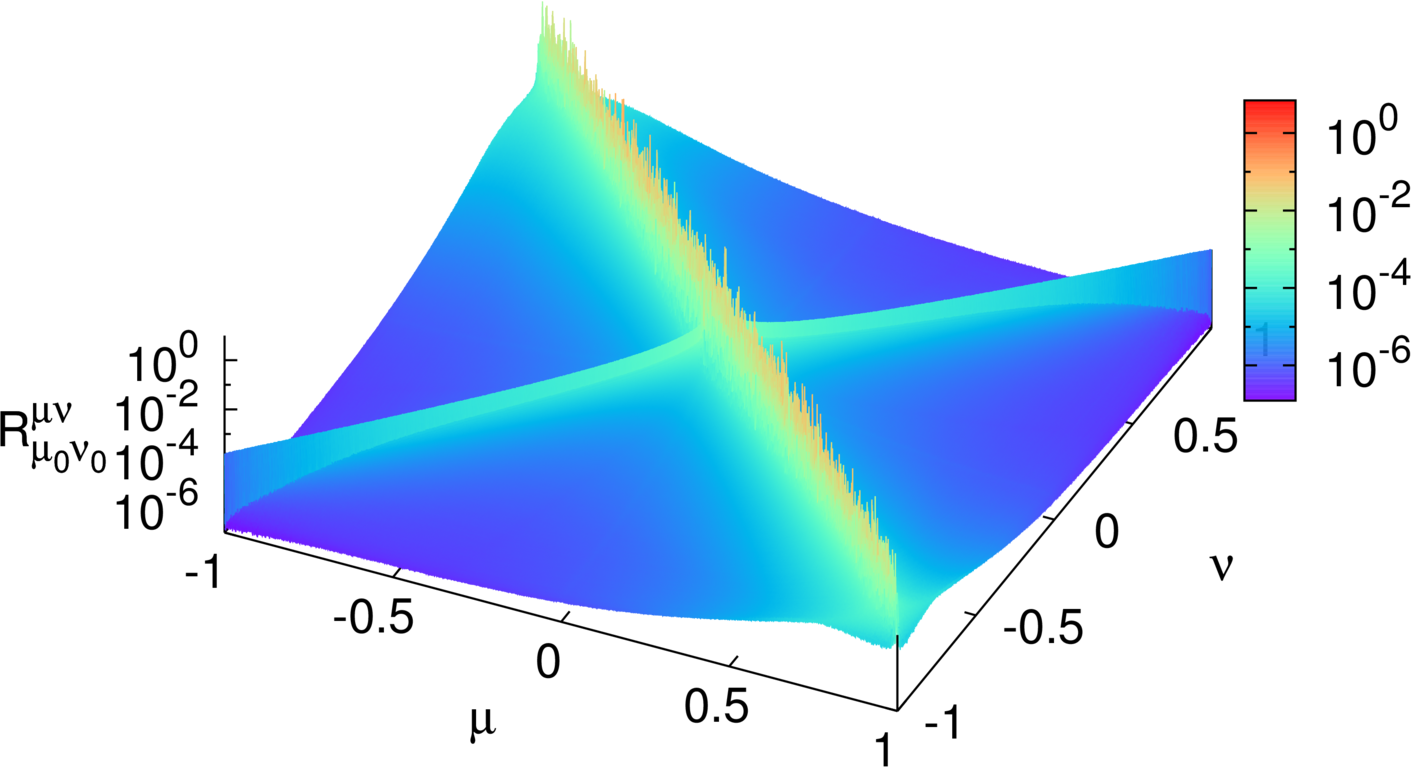}
    \par\end{centering}}
    \subfloat[]{
    \begin{centering}
        \includegraphics[width=0.33\textwidth]{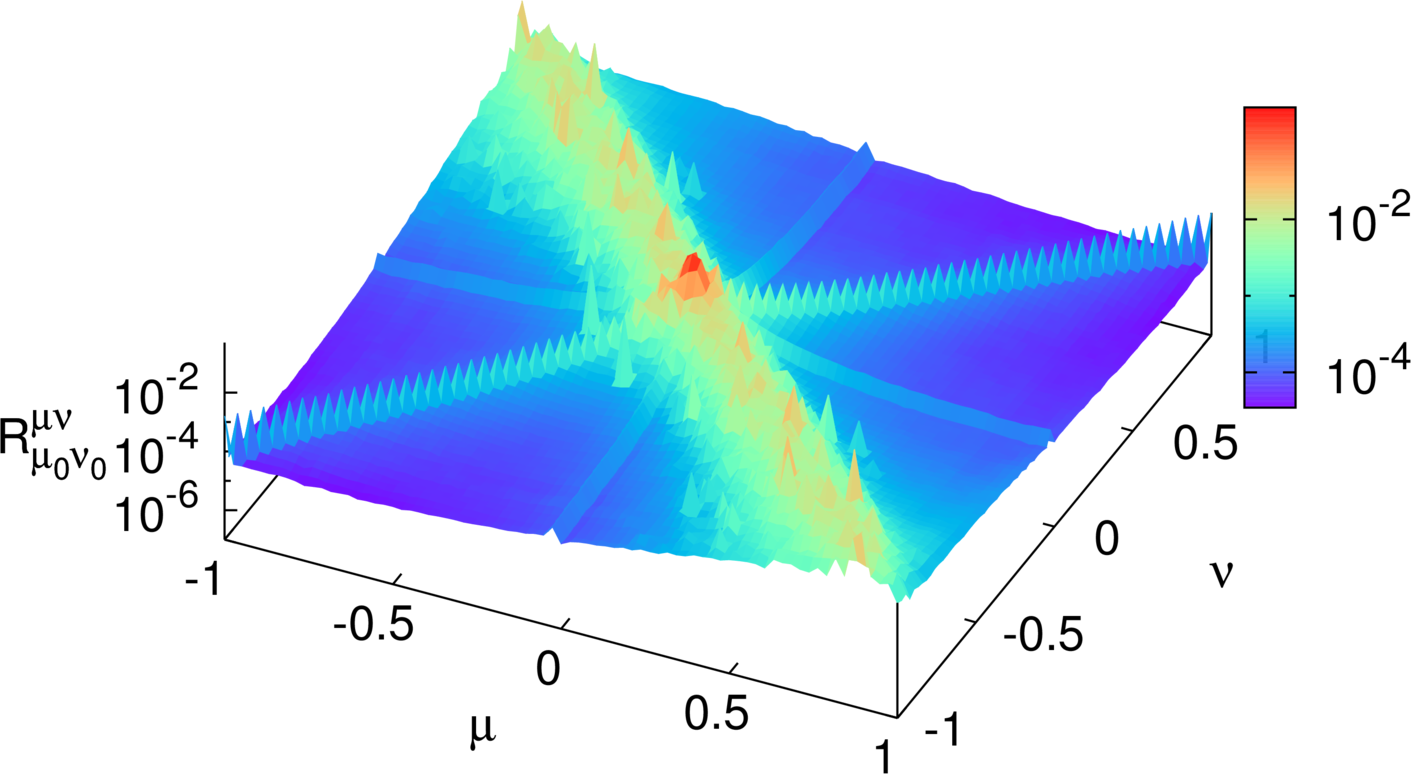}
    \par\end{centering}}
    \subfloat[]{
    \begin{centering}
        \includegraphics[width=0.33\textwidth]{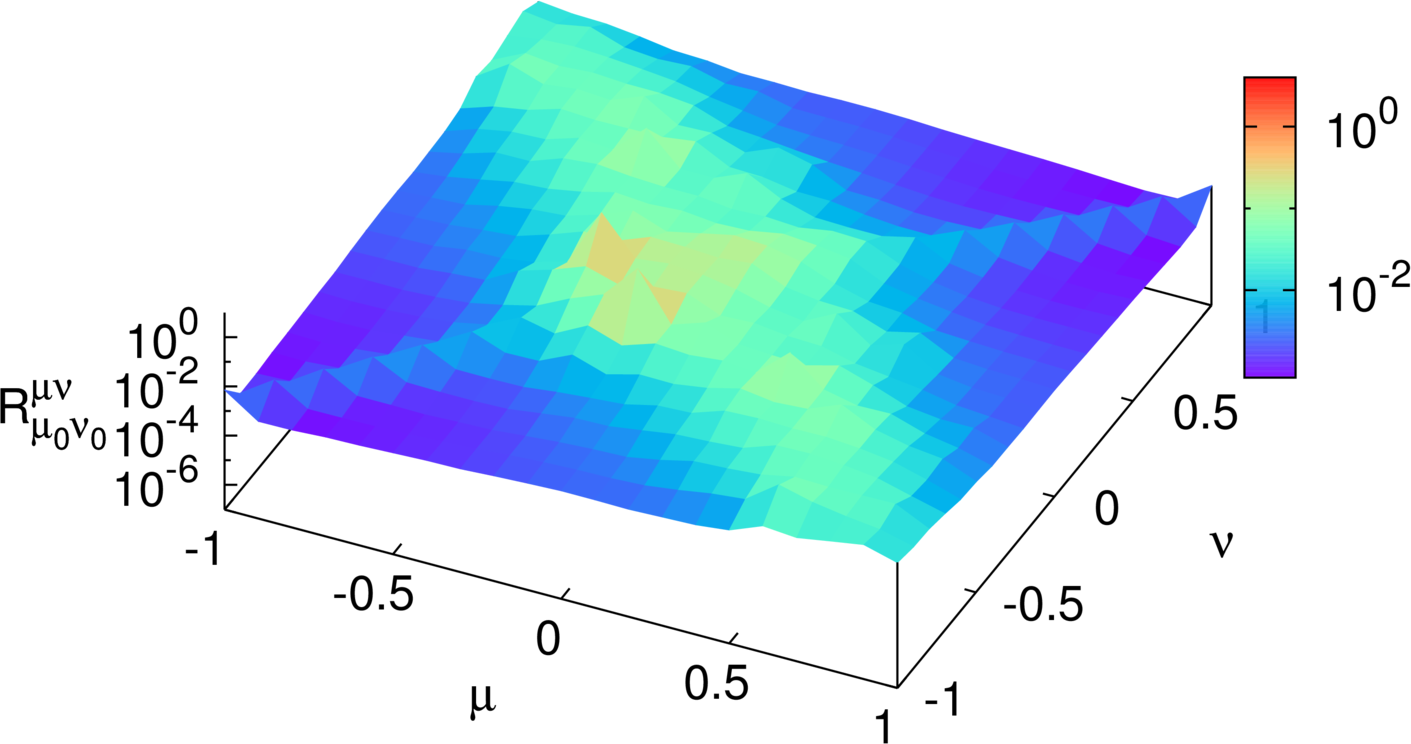}
    \par\end{centering}}
    \caption{(Color online) Coupling strength ratio $R$ for $\mu_0=\nu_0$ and $E_{\mu_0} \approx 0$ for $u=1.0$ versus $\nu,\mu$ in units of $\xi_1$. (a) $W=0.5$, (b) $W=2$ and (c) $W=4$. Data are averaged over $5000$ disorder realizations.}
    \label{fig:couplingStrength}
\end{figure*}

\section{Model}

The single particle Anderson Hamiltonian in one space dimension is given by
\begin{gather}
    \label{eq:ham-1p}
    H_0 = \sum_n \left[t(|n\rangle\langle n+1| + |n+1\rangle\langle n|) + V_n|n\rangle\langle n|\right],
\end{gather}
where $|n\rangle$ denotes a basis state with one particle located on site $n$, $t$ denotes the nearest neighbor hopping which we fix to unity in all computations $t=1$, and $V_n$ is an onsite potential, sampled from a uniform distribution $[-W/2,W/2]$. $W$ characterizes the strength of disorder in the system. The eigenenergies $\{E_\mu\}\in[-2t-\frac{W}{2},2t+\frac{W}{2}]$ and the eigenfunctions $\{\phi_{\mu}(n) \equiv |\mu\rangle\}$ of the single particle problem~\eqref{eq:ham-1p} are obtained by diagonalising $H_0$. The spectrum is symmetric around $E=0$ and has width $\Delta_1=4t+W$. The single particle localization length $\xi_1$ controls the exponential decay of an eigenfunction $\phi_{\mu}(|n| \rightarrow \infty ) \sim {\rm e}^{-|n|/\xi_1}$. For $W < 4$ the largest localization length $\xi_1$ at $E=0$ is well approximated by $\xi_1 \approx 100 t^2/W^2$. 

Two indistinguishable particles are described with the basis states $|n,m\rangle$, where $n$ and $m \geq n$ stand for their coordinates. The two interacting particles (TIP) Hamiltonian is then defined with the use of  the basis states $|n,m\rangle$ as:

\begin{gather}
    \label{Ham}
    H = H_0\oplus H_0 + uP \;.
\end{gather}

The operator $P\ket{n,m}=\delta_{nm}\ket{n,m}$ projects a two particle state onto the basis states with doubly occupied sites. The parameter $u$ controls the strength of the interaction. Since we address states with the largest localization length in the center of the spectrum with only two particles involved, neither the type of interaction (repulsive with positive $u$ or attractive with negative $u$) nor the particle statistics will influence the quest for the asymptotic scaling. We use bosons for convenience.

\section{Fock space connectivity}

We first evaluate the interaction-induced connectivity in Fock space in order to establish the asymptotic scaling regime parameters, and in order to estimate the relevant energy scales. A Fock state $\left(\mu,\nu\right)\equiv |\mu\rangle \otimes  |\nu\rangle$  is an eigenstate of the two particle system for $u=0$ with eigenenergy $E_{\mu,\nu} \equiv E_{\mu} + E_{\nu}$. The interaction induces a matrix element $\langle \mu,\nu | u P |\mu^{\prime},\nu^{\prime} \rangle = u I_{\mu,\nu}^{\mu^{\prime},\nu^{\prime}}$ between two Fock states with the overlap integral
\begin{gather}
    \label{eq:overlapIntegrals}
    I_{\mu,\nu}^{\mu^{\prime},\nu^{\prime}}= \sum_{n}\phi^*_{\mu^{\prime}}(n) \phi^*_{\nu^{\prime}}(n) \phi_{\mu}(n)\phi_{\nu}(n) \; .
\end{gather}
The energy difference between the chosen two Fock states is given by $\Delta E = E_{\mu '}+E_{\nu '}-E_{\mu}-E_{\nu}$. A strongly connected pair of Fock states is found if the ratio~\cite{krimer2015interaction}
\begin{gather}
    \label{eq:couplingStrength}
    R_{\mu,\nu}^{\mu ',\nu '}\equiv\left|\frac{u I_{\mu,\nu}^{\mu^{\prime},\nu^{\prime}}}{E_{\mu '} + E_{\nu '} - E_{\mu} -E_{\nu}}\right|>1 \; .
\end{gather}
Krimer et al showed in Ref.~\onlinecite{krimer2015interaction} that the overlap integrals turn from random like for $W > 1$ to selective ones in accord with the restoration of translational invariance and the corresponding momentum conservation for $W < 1$, thus invalidating the analytical considerations in Refs.~\onlinecite{shepelyansky1994coherent,imry1995coherent} in accord with earlier predictions by Ponomarev et al.~\cite{ponomarev1997coherent} 

Here we follow the computational method of Ref.~\onlinecite{krimer2015interaction}. In a nutshell, we choose a reference single particle state $|\mu_0\rangle$ with $E_{\mu_0} \approx 0$ and reference Fock state $|\mu_0,\nu_0\equiv \mu_0 \rangle$. We find all single particle states $\left \{ |\nu\rangle \right \}$ whose coordinates $x_{\nu} = \sum_n n \phi_{\nu}(n)$ are within the range of one localization length distance from $x_{\mu_0}$, i.e. $|x_{\nu} - x_{\mu_0}| \leq \xi_1$. Each set is sorted with ascending energy $E_{\nu}$ with the convention $\nu_0 = 0$. In the limit of weak disorder the index $\nu$ will be related to a wave number. Each pair $|\mu,\nu\rangle$ corresponds to a new Fock state which may have significant overlap $R > 1$ with the reference state. The resulting disorder averaged (5000 realizations) distribution of $R$ is plotted versus $\nu/\xi_1,\mu/\xi_1$ in Fig.\ref{fig:couplingStrength} for three different values of $W=0.5,2,4$. The broad distribution for $W=4$ in Fig.~\ref{fig:couplingStrength}(c) is replaced by one major thin resonance line $\nu = \mu$ for $W=0.5$ in Fig.~\ref{fig:couplingStrength}(a). The restoring of translational invariance ~\cite{frahm2016eigenfunction,jacquod1997breit} leads to a momentum conservation related selection rule of plane waves with fixed boundary conditions~\cite{krimer2015interaction} and approximately reads $\nu_0 + \mu_0 = \nu + \mu$ which results in $\nu = - \mu$ with our reference state choice.

In addition the largest coupling strength ratio $R$ is obtained for the smallest energy differences $\Delta E$ which  imply $E_{\mu} = -E_{\nu}$, resulting again in the condition $\nu = -\mu$ due to the particle-hole symmetry of the single particle spectrum. As a result, for weak disorder $W < 1$ the asymptotic connectivity regime of Fock states is observed. In this regime, energy and momentum are approximately conserved, and only a selected set of about $\xi_1$ Fock states with $|\nu,-\nu\rangle$ is strongly connected. Their overlap integrals can be approximated replacing $\phi_{\nu}(n) \approx {\rm e}^{i\nu n}/\sqrt{\xi_1}$ by plane waves normalized to a box of size $\xi_1$. It follows $I \approx 1/\xi_1$. The Fock state energies of that group are confined to an interval of width $\Delta_1/\xi_1$ due to energy conservation, and result in a level spacing $\delta_2 \approx \Delta_1 /  \xi_1^2$. We conclude that the selected set of strongly interacting Fock states is characterized by an effective disorder $W_{eff} \approx \Delta_1 /  \xi_1^2$ and an effective hopping $t_{eff} \approx u/\xi_1$. A naive use of the localization length estimate for a corresponding tight binding chain would result in $\xi_2 \sim 100 \frac{u^2}{\Delta_1^2} \xi_1^2$.

\begin{figure}[hbt]
    \includegraphics[width=0.49\textwidth,angle=0]{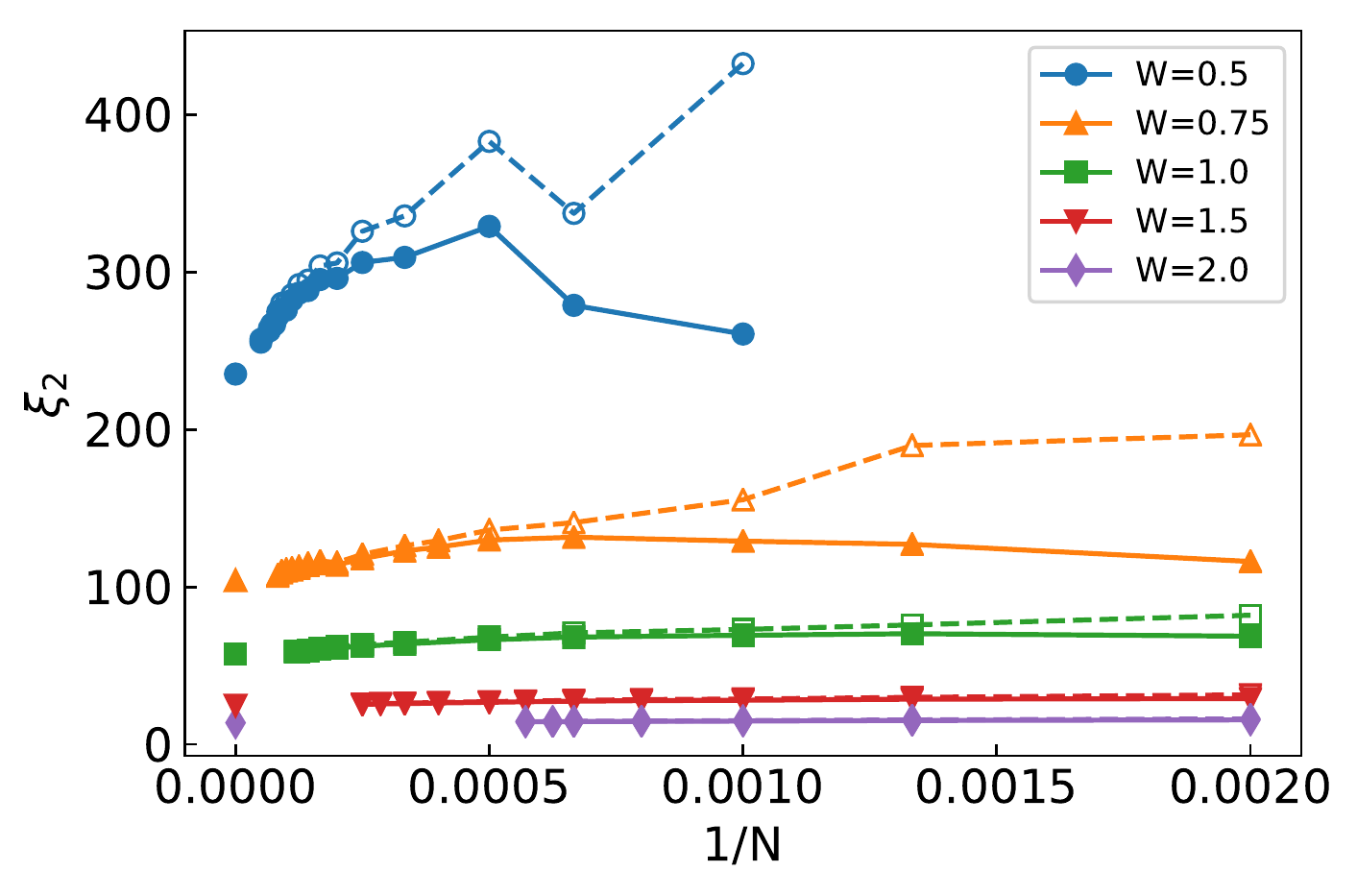}
    \caption{(Color online) Plot of $\xi_2$ vs $1/N$ for $u=0$ and various values of $W$ obtained by the two methods (see the text): M1 (filled symbols and solid lines) and M2 (empty symbols and dashed lines). The corresponding extrapolated values of $\xi_2(N\to\infty)$ are located at $1/N=0$.}
    \label{fig:2}
\end{figure} 

\begin{figure}[h]
    \centering
    \includegraphics[width=0.49\textwidth,angle=0]{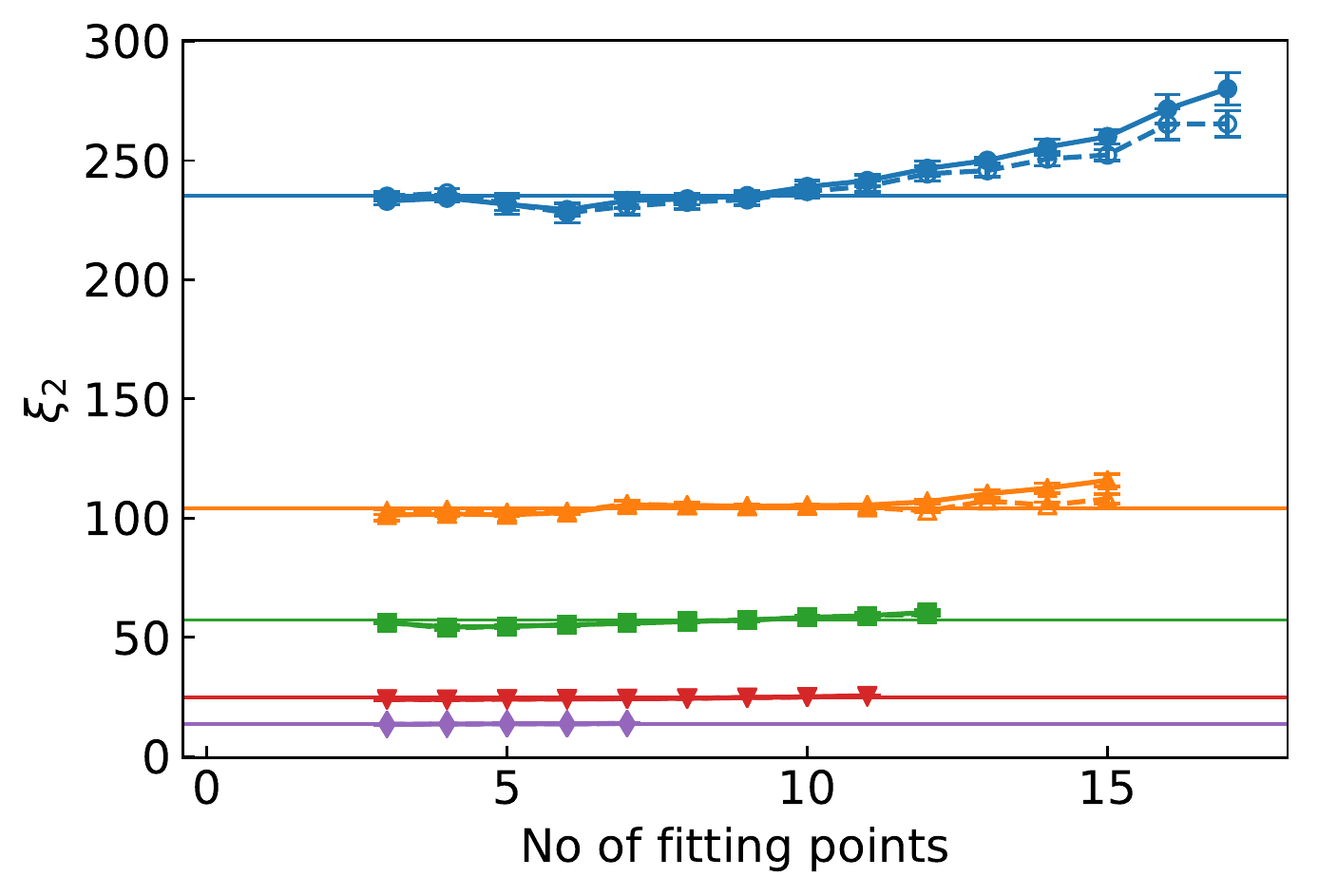}
    \caption{(Color online) The extrapolated value $\xi_2(1/N \rightarrow \infty)$ with error bars (some error bar values smaller than the legend size are not visible) versus the number of data points from Fig.~\ref{fig:2} ($u=0$) counted from the largest towards the smallest system sizes. The horizontal lines show the values extracted and used for the forthcoming analysis. Sufficiently large system sizes $N > N_*(W)$ are required to see the convergence.}
    \label{fig:2b}
\end{figure} 

\begin{figure*}[bth]
    \centering{}\subfloat[]{
    \begin{centering}
        \includegraphics[width=0.33\textwidth]{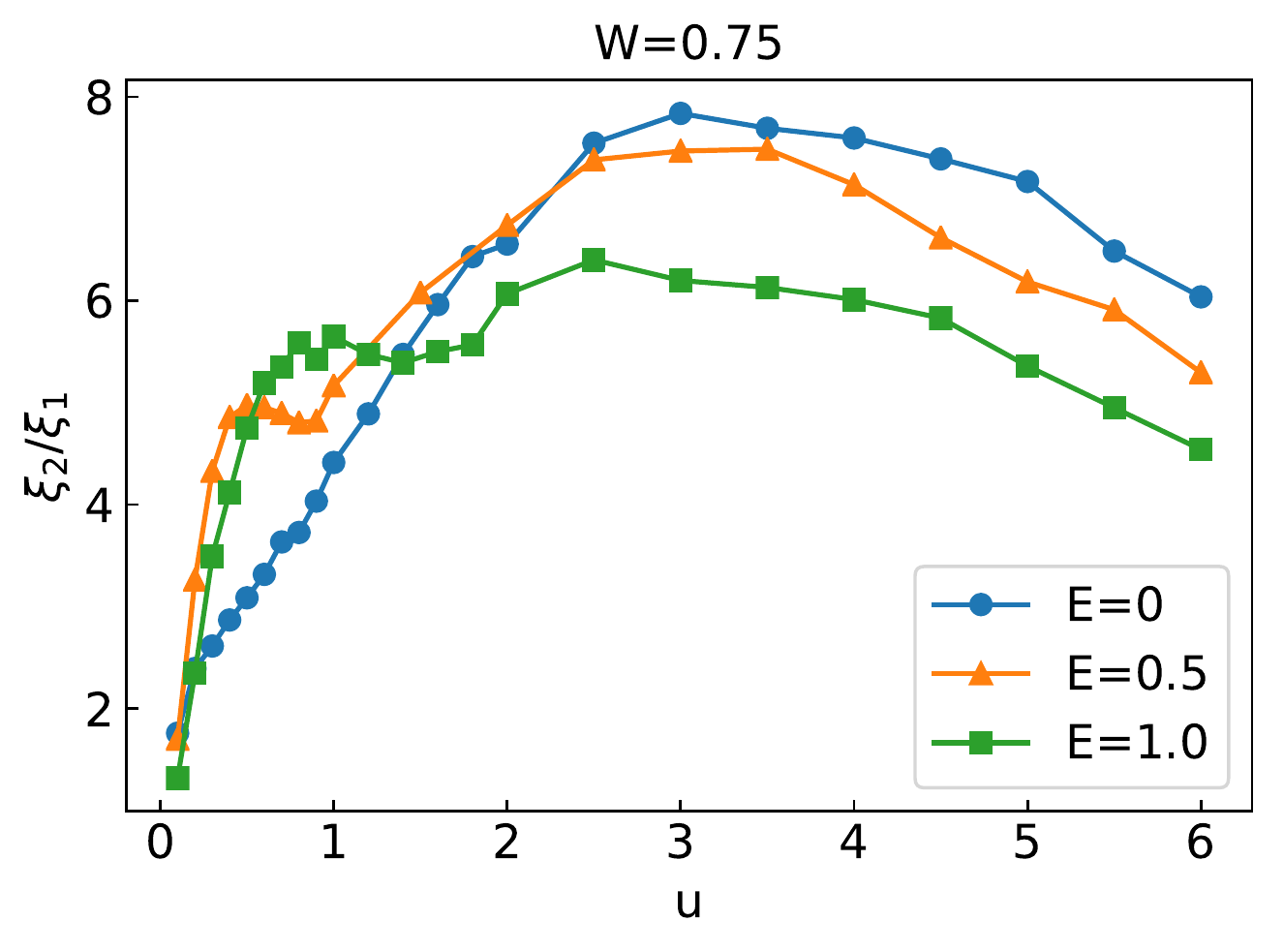}
    \par\end{centering}}
    \subfloat[]{\begin{centering}
        \includegraphics[width=0.33\textwidth]{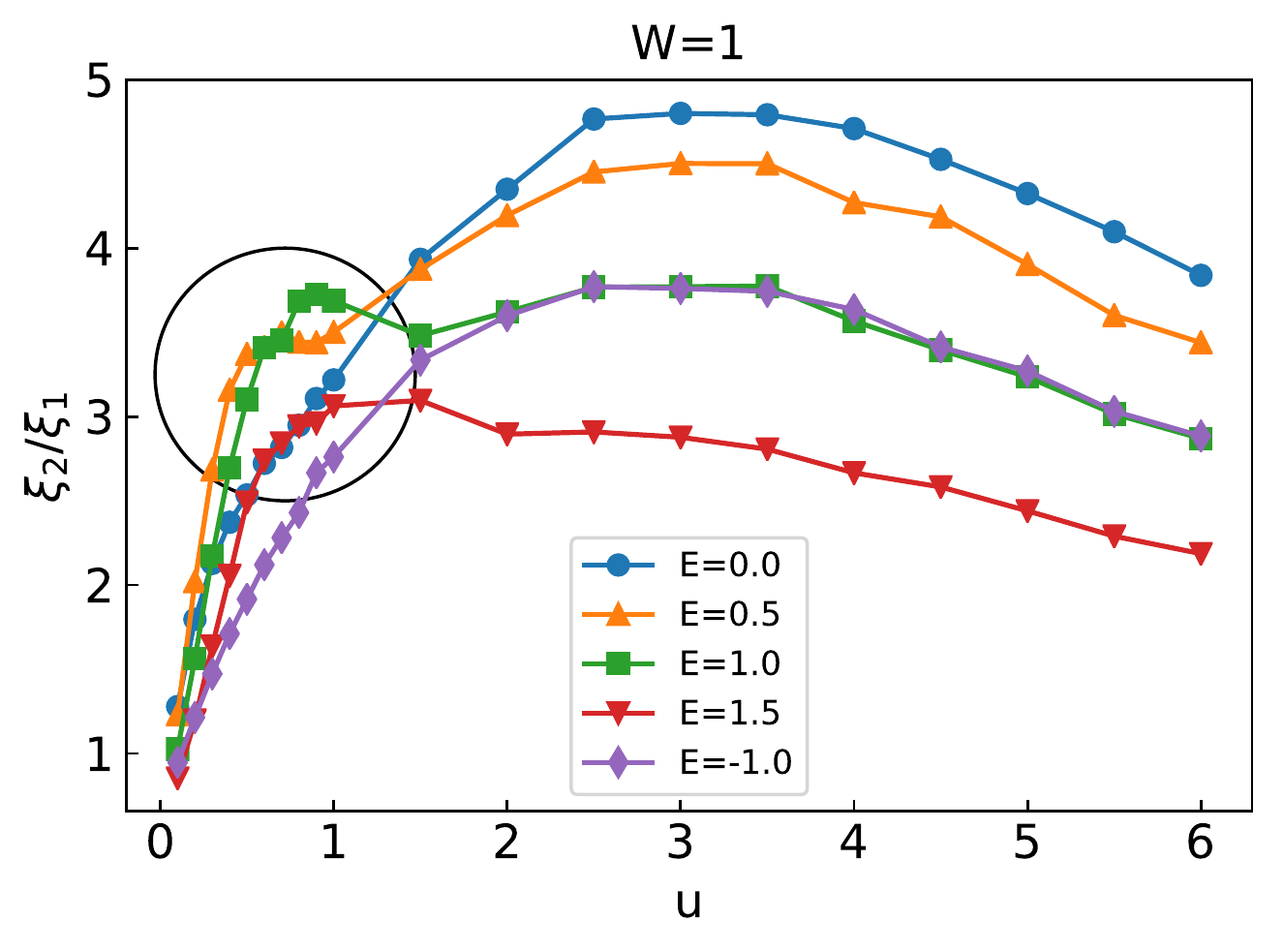}
    \par\end{centering}}
    \subfloat[]{\begin{centering}
        \includegraphics[width=0.33\textwidth]{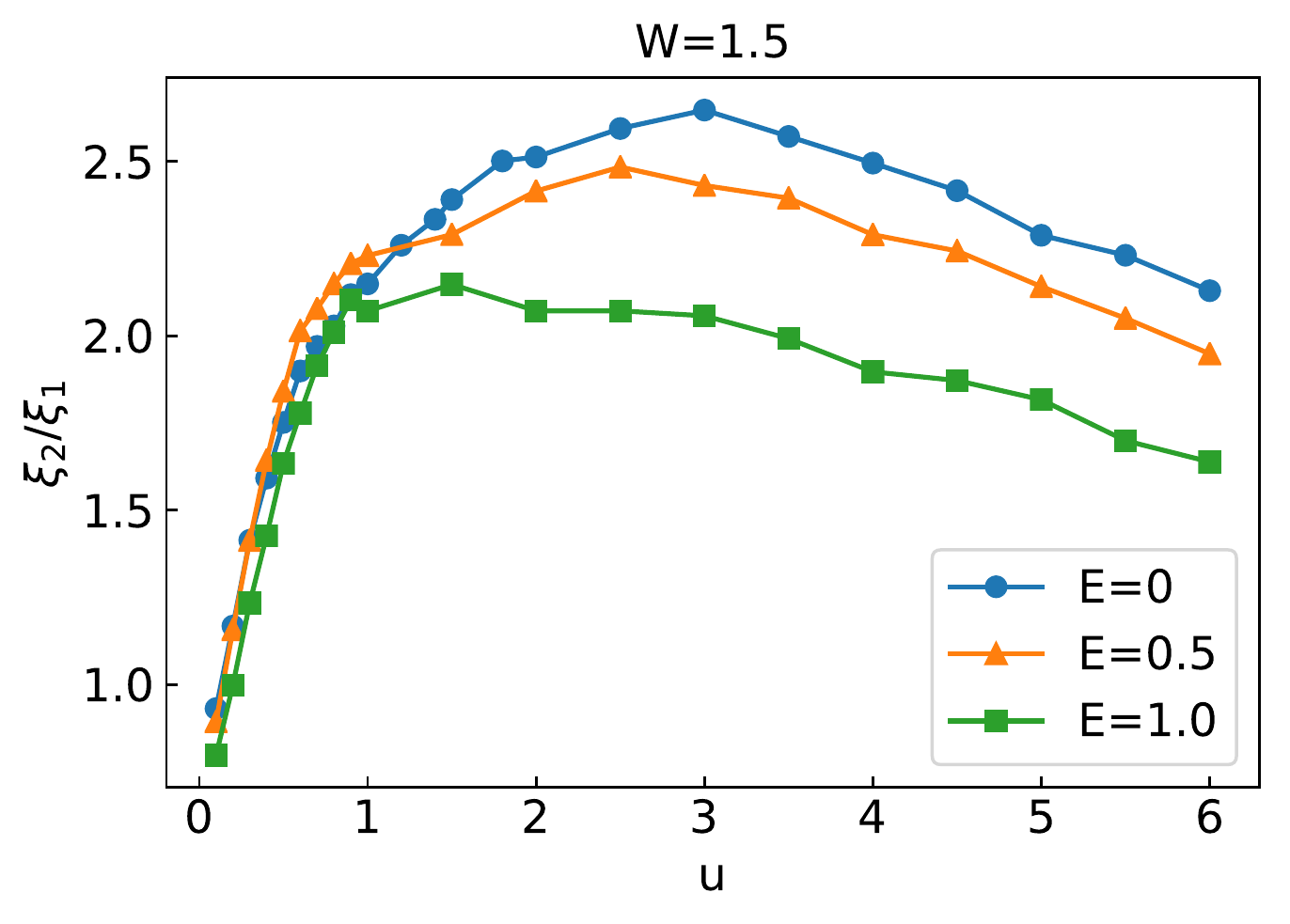}
    \par\end{centering}}
    \caption{(Color online) $\xi_2/\xi_1$ vs $u$ for various energies $E$. a) $W=0.75$ and $E=0,0.5,1$; b) $W=1$ and  $E=0, 0.5, 1.0, 1.5, -1.0$;  c) $W=1.5$ and $E=0,0.5,1$. The circle in plot b) shows the location of an anomaly where $\xi_2$ at a given value of $u$ is enhanced for non-zero energy values $E$.}
    \label{fig:3}
\end{figure*}

\section{Measuring $\xi_2$}
\label{sec:xi2}

\subsection{Green function method: Benchmarking the non-interacting case}
\label{sec:xi2u0}

To proceed we compute the localization length $\xi_2$, following Ref.~\onlinecite{vonoppen1996interaction} where it was shown that the full Dyson equation for the two-particle Green's function (GF) $G=(E-H)^{-1}$ can be solved for basis states with double occupancy using the non-interacting $u=0$ GF $G_0=(E-H_0)^{-1}$ as
\begin{gather}
    \tG = \frac{\tG_0}{1 - u\tG_0},
\end{gather}
where $\tG = PGP$ and $\tG_0 = PG_0P$ are projections of the full and non-interacting GFs onto double occupancy sites. This remarkable result allows to assess the wave function of the interacting system on double occupied sites through solving the non-interacting eigenvalue problem, and thus allows to measure the TIP localization length enhancement. The localization length $\xi_2$ is then defined as the exponent of the exponential decay of $\tG$:
\begin{gather}
    \label{eq:xi2-def}
    \frac{1}{\xi_2} = -\lim_{|n-m| \to\infty} \frac{\overline{\ln|\langle n,n|\tilde{ G} |m,m\rangle|}}{|n-m|}
\end{gather}
where $\overline{...}$ denotes the disorder average. 
To obtain the projected GF $\tG_0$ we solve the single particle eigenvalue problem and compute
\begin{gather}
\label{eq:G0}
    \langle n,n| G_0(E)|m,m\rangle = \sum_{\mu,\nu} \frac{\phi_\mu(n)\phi_\nu(n)\phi_\mu^*(m)\phi_\nu^*(m)}{E - E_\mu - E_\nu}.
\end{gather}

The complexity of this expression if $O(N^4)$ which is prohibitive for large sizes $N$ that we seek to explore. It was shown in Ref.~\onlinecite{frahm1999interaction} that this complexity reduces to $O(N^3)$ if one exploites the tridiagonal structure of the single particle Hamiltonian~\eqref{eq:ham-1p}. Namely one reorders the summation in Eq.~\eqref{eq:G0} as 
\begin{gather}
    \langle n,n| G_0(E)|m,m\rangle = \sum_{\nu} \phi_\nu(n) g_{nm}(E-E_\nu) \phi_\nu(m)\;,\\
    g_{nm}(E) = \sum_\mu \frac{\phi_\mu(n) \phi_\mu(m)}{E-E_\mu}\;.
\end{gather}
where the single particle Green function $g_{nm}(E - E_\mu)$ is evaluated using a fast dedicated inversion algorithm for tridiagonal matrices.

The subsequent processing of the data is organised in two steps:

Step 1: disorder averaging and real space fitting. For a given system size $N$ we use GF data for $0.05 N \leq n,m \leq 0.95 N$ to avoid possible boundary effects and define a function $y(x) \equiv \ln|\langle n,n| \tilde{G} |m,m\rangle|$ where $x\equiv |n-m|$. In method M1 we compute the disorder averaged $\overline{y(x)}$ and then find the best linear fit $a-x/\xi_2$ of the TIP localization length $\xi_2(N)$. In method M2 we first fit $y(x)$ with a linear function $a-x/\xi$ and compute $\xi_2(N)=\overline{\xi}$ as the disorder average of $\xi$. For a given system size $N$, the number of disorder realisations $S$ is chosen such that $NS\approx10^6$ with $N$ going up to $N=20000$.


We benchmark both methods for $u=0$. Figure~\ref{fig:2} shows $\xi_2(N)$ for $W=0.5,0.75,1.0,1.5,2$ at $u=0$ obtained from both methods M1 and M2. We find that both methods agree once $N \geq N_*(W)$ with $N_*(W) \approx 6000, 2000, 1500, 750, 500$ for $W = 0.5, 0.75, 1, 1.5, 2$ respectively. This implies a power-law dependence $N_*(W)\propto W^{-1.72}$. For larger disorder, $W>2$, the value of $N_*(W)$ is effectively zero for any reasonable system size. Importantly, we conclude that we need system size $N > 2000$ for disorder strengths $W=0.75, 0.5$ for the outcomes of both methods to coincide. These system sizes were not addressed in the literature before.
\\
Step 2: finite size fitting for $N \geq N_*(W)$. Using $\xi_2(N) = \xi_ 2 + a/N$ (see e.g. Ref.~\onlinecite{song1997localization}) we extract $\xi_2(N\rightarrow \infty)$ using data from the both methods M1, M2. The results are presented in Fig.~\ref{fig:2b}. We observe that the fitting is insensitive to the choice of the methods M1 and M2 as well as to the number of data points (system sizes $N$) used to do the fitting as long as $N \geq N_*(W)$. Therefore we use the M1 method in the subsequent analysis for non-zero $u$. Further details about fits and extrapolations are discussed in the Appendix.


According to Ref.~\onlinecite{vonoppen1996interaction} $\xi_2/\xi_1=0.5$ at $u=0$. Our data yield up to $10\%$ error with $\xi_2/\xi_1 \approx 0.56$ for the weakest disorder $W=0.5$. This has to be compared against the $30\%$ error obtained for smaller system sizes in Ref.~\onlinecite{frahm2016eigenfunction} ($\xi_2 / \xi_1 \approx 0.65$ for $W=0.5$).

\subsection{Non-zero interactions}

\begin{figure*}[t]
    \centering{}\subfloat[]{
    \begin{centering}
        \includegraphics[width=0.43\textwidth]{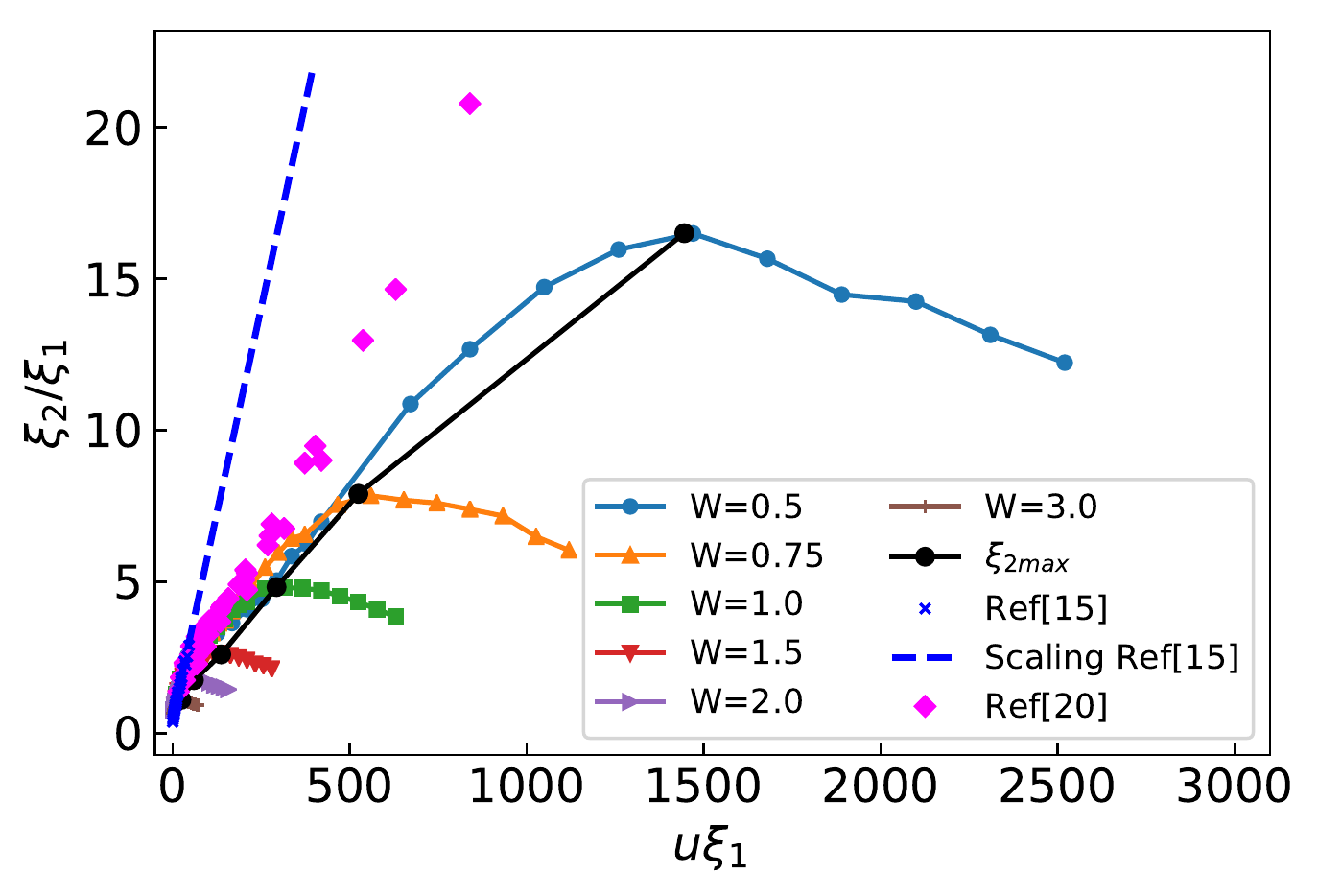}
    \par\end{centering}}
    \subfloat[]{\begin{centering}
        \includegraphics[width=0.43\textwidth]{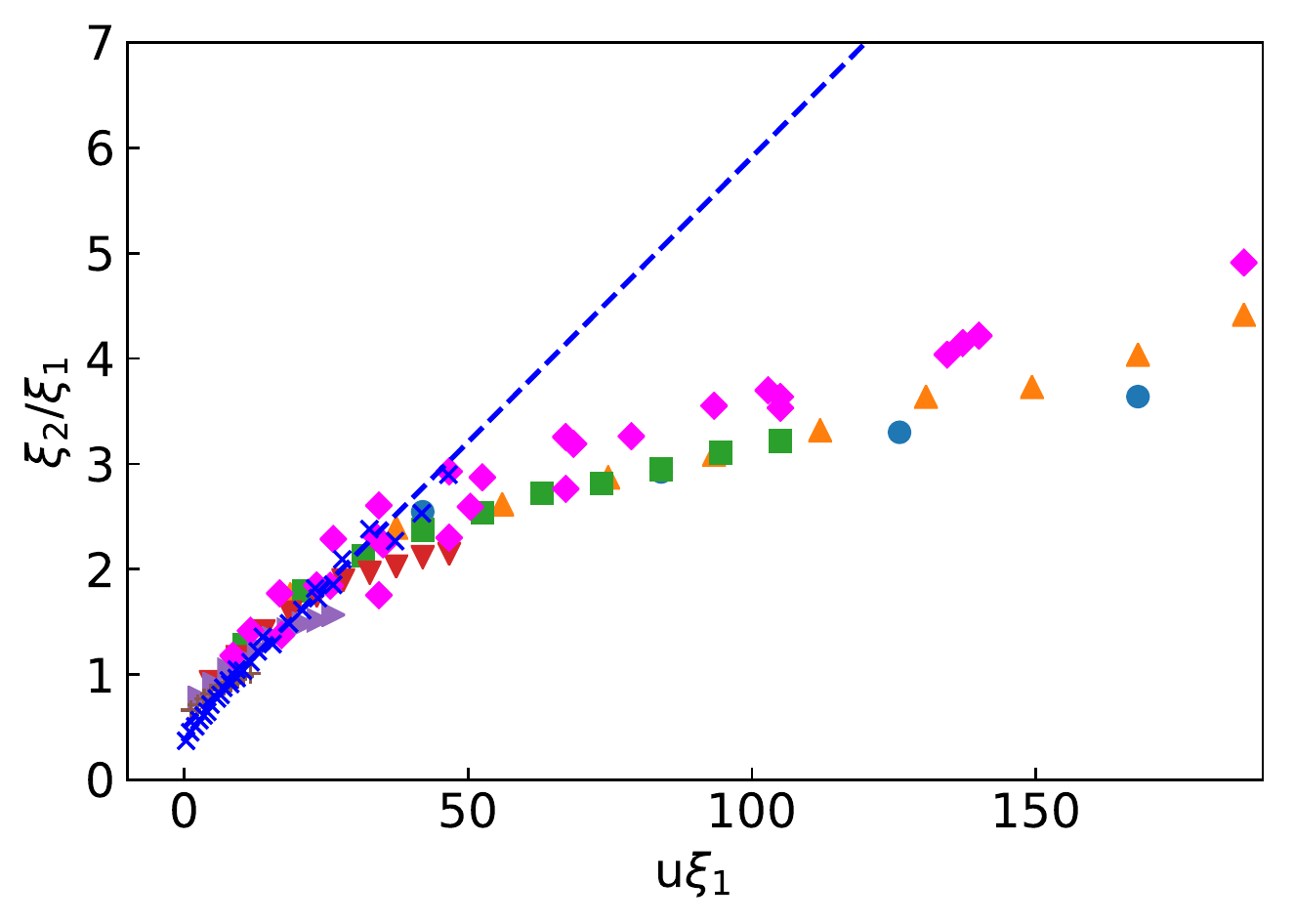}
    \par\end{centering}}
    \caption{(Color online) (a) $\xi_2/\xi_1$ versus $u\xi_1$ for different values of $W$. The largest obtained values for $\xi_2$ are connected with a solid black line. Data from Ref.~\onlinecite{vonoppen1996interaction} for $W=1.5,2,3,4,5$ and Ref. \onlinecite{frahm2016eigenfunction} for $W=0.5,0.625,0.75,0.875,1,1.5,1.75$ are included for comparison. The scaling result from Ref.~\onlinecite{vonoppen1996interaction} is shown with a dashed line. (b) A zoom of the universal part from (a) e.g. up to the maximum $\xi_2/\xi_1$ to highlight the crossover into the asymptotic scaling region at $u\xi_1 \approx 50$.}
 \label{fig:4}
\end{figure*}

To proceed to nonzero interactions, we compute $\xi_2/\xi_1$ as a function of $u$ for different values of the energy $E$ and disorder strength $W$ as shown in Fig.~\ref{fig:3}. For fixed $E$ and $W$ we find that $\xi_2$ passes through a maximum $u_{max}$ upon increasing $u$ and decreases with further increasing $u$. This decreasing for $u \geq \Delta_1$ is due to doubly occupied site states being tuned out of the spectrum of the remaining two-particle continuum, see Ref.~\onlinecite{ponomarev1997coherent} for details. The value of $u_{max} \approx 3t$ is in agreement with the prediction from Ref.~\onlinecite{ponomarev1997coherent} as well. In addition we also observe an anomaly~\cite{frahm2016eigenfunction} at $u \approx t$ and $E \approx 1$ for $W \leq 1$ where the localization length $\xi_2$ is enhanced as compared to $E=0$. However, the record value of $\xi_2$ for each studied disorder strength is found for the band center $E=0$ and the interaction strength $u_{max} \approx 3t$. As shown above, the resonantly interacting Fock state groups contain $\sim \xi_1$ member states and are characterized by two energy scales - an effective disorder strength  $W_{eff} \approx \Delta_1 /  \xi_1^2$ and an effective matrix element $t_{eff} \approx u/\xi_1$. The ratio of both yields a dimensionless new parameter $t_{eff}/W_{eff} = u\xi_1/\Delta_1$. In the limit of weak disorder $\Delta_1$ turns simply into a constant leaving us with the relevant parameter $u \xi_1$. We then plot $\xi_2/\xi_1 \equiv F(u \xi_1)$ versus $u\xi_1$ in Fig.~\ref{fig:4}(a). For $u\xi_1 < 50$ we find agreement with the data from~\cite{vonoppen1996interaction} which indicate $F(x)\sim x$. Lowering $W < 1$ leads to an increase of $u\xi_1$ and a consequent crossover into the asymptotic scaling regime, which shows a significant slowing down of the increase of the TIP localization length. The corresponding scaling function $F(x)$ turns nonlinear with sublinear growth for large values of $x > 50$. The extrapolation of the data from~\cite{vonoppen1996interaction} (dashed line) overestimates the TIP localization length by at least a factor of 6 for values $x=u\xi_1 = 1500$. Similarly the data from Ref.~\onlinecite{frahm2016eigenfunction} overestimate the length by a factor of up to 2 for $x=1500$, probably due to finite size effects which we did take into account. The crossover  into the asymptotic regime at $x \approx 50$ is highlighted in Fig.~\ref{fig:4}(b). Notably the record values of the TIP localization length are obtained for $u\approx 3$. The solid line in Fig.~\ref{fig:4}(a) connects these data points, and indicates sublinear growth of $\xi_2/\xi_1$ with $u\xi_1$ and a corresponding nonlinear scaling function $F(x)$. 

\begin{figure}[h]
    \centering
    \includegraphics[width=0.45\textwidth,angle=0]{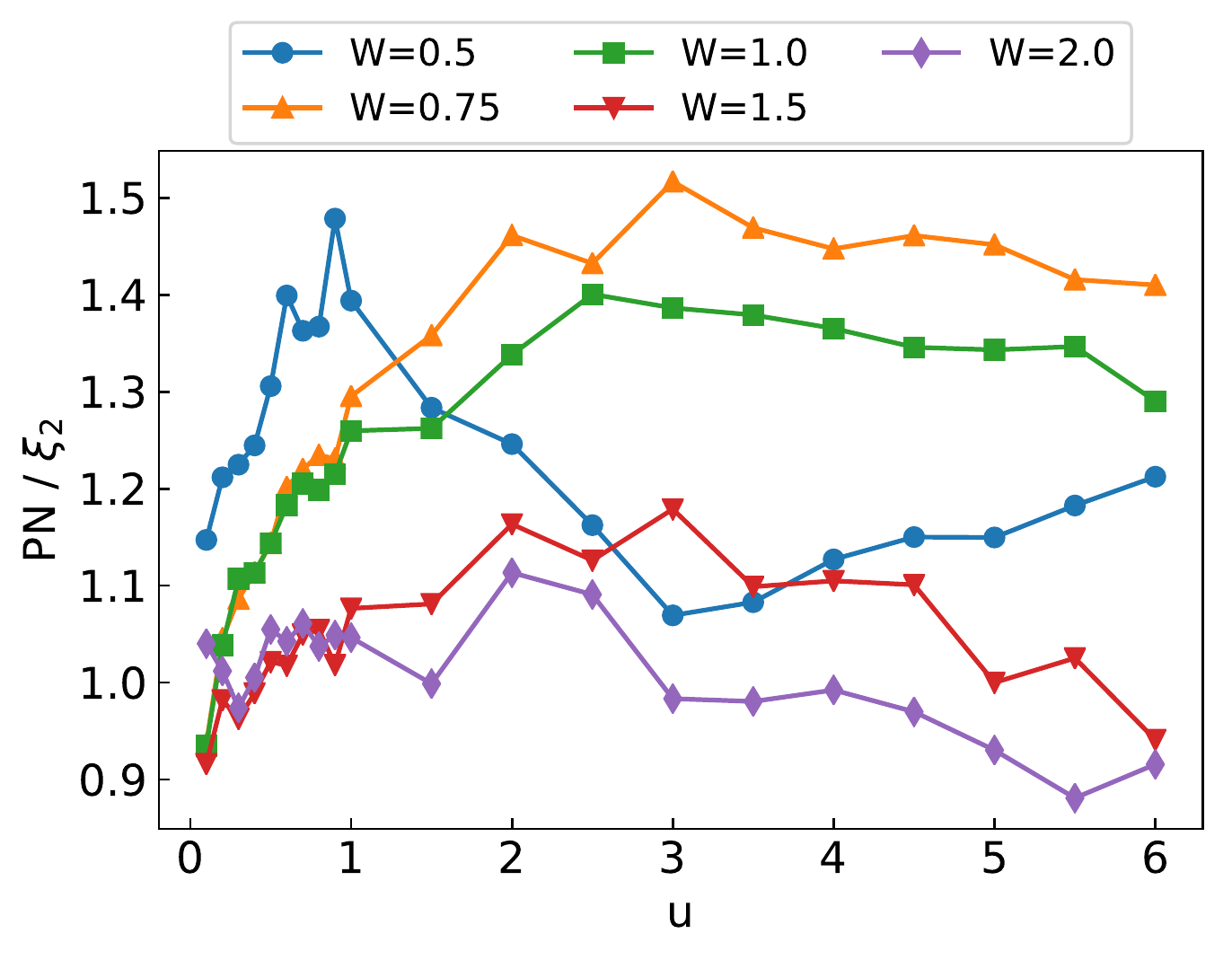}
    \caption{(Color online) Participation number $\pn$ in units of the computed localization length $\xi_2$ as a function of $u$ for various values of disorder $W$.} 
    \label{fig:5}
\end{figure}

In order to test the single parameter scaling hypothesis,~\cite{kramer1993localization} we compute the participation number as 
\begin{gather}
\pn = \frac{\xi_2}{N} \sum_{l=1}^{N/\xi_2} P(n\equiv l \xi_2)\;,\;
P(n) = \frac{(\sum_m |\tG_{n,m}|)^2 }{  \sum_m|\tG_{n,m}|^2}\;.
\end{gather}
As before the number of disorder realizations $S$ for a given system size $N$ was fixed such that $NS\approx 10^6$. Additionally we perform a finite size scaling to eliminate finite size corrections. The values of the ratio $\pn/\xi_2$ for $W=0.5, 0.75, 1.0, 1.5, 2.0$ are plotted in Fig.~\ref{fig:5}. We observe that $1 \leq \pn/\xi_2  \leq 1.5$ which indicates that the extension of the TIP wavefunctions is of the same order as the localization length $\xi_2$ which controls their exponential decay, confirming the single parameter scaling.

\section{Discussion}

In conclusion, we showed that two interacting particles in a disordered potential enter an asymptotic scaling regime of their localization length $\xi_2$ in units of the single particle localization length $\xi_1$ for weak disorder due to the restoring of momentum conservation in the single particle eigenfunctions. 
The ratio $\xi_2/\xi_1= F(u \xi_1)$ grows to record values of $F = 15$ for $\xi_1 = 400$ and $u=3$, albeit the growth is much slower than anticipated from earlier numerical studies and reflected in a nonlinear dependence of the scaling function $F(x)$ on its argument. Our findings are supported by the manifestation of resonant couplings between fragile groups of Fock states which have much smaller size ($\xi_1$) than originally anticipated ($\xi_1^2$). Further, the fragility is supported by the presence of particle-hole symmetry due to the bipartite nature of the tight binding chain for a single particle in the limit of weak disorder. This particle-hole symmetry guarantees that resonant groups of Fock states can conserve both momentum and energy. We expect that a violation of particle-hole symmetry e.g. by adding next-to-nearest-neighbour hoppings will reduce the size of the resonantly interacting Fock state groups and further reduce the enhancement factor of $\xi_2$ over $\xi_1$. The nature of the observed non-linear correction of the scaling function $F(x)$ is an interesting subject of future studies.

\begin{acknowledgments}
    We thank Boris Altshuler and Mikhail Fistul for stimulating discussions. This work was supported by the Institute for Basic Science in Korea (IBS-R024-D1). T.E. acknowledges financial support by the Alexander-von-Humboldt foundation through the Feodor-Lynen Research Fellowship Program No. NZL-1007394-FLF-P.
\end{acknowledgments}

\appendix

\renewcommand\thefigure{\thesection.\arabic{figure}} 
\setcounter{figure}{0}

\bigskip

\section{Fitting and extrapolating}

\begin{figure}[h]
    \centering
    \includegraphics[width=0.49\textwidth,angle=0]{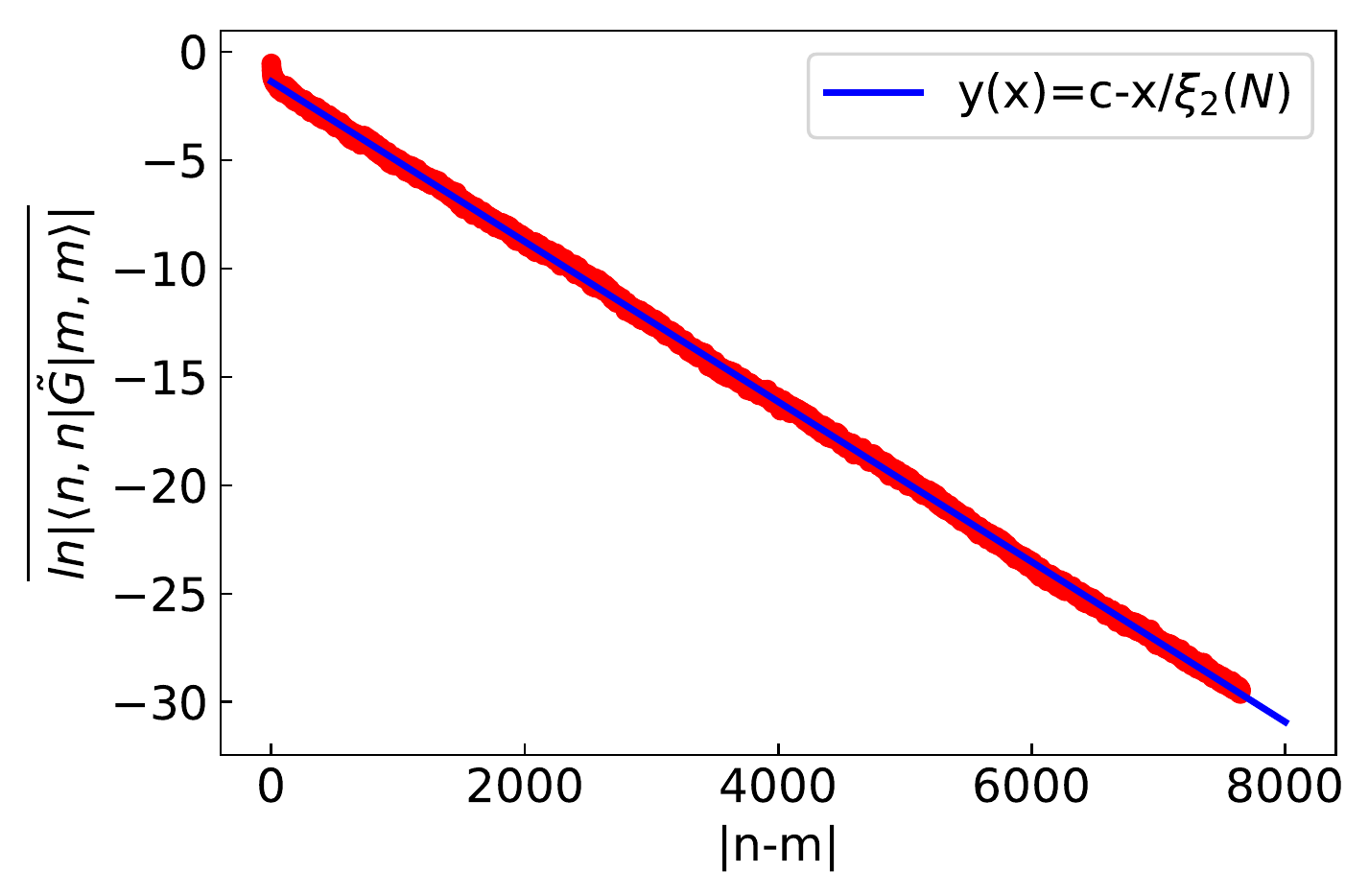}
    \caption{(Color online) Decay of $y(x)= ln\overline{\langle n,n|\tG|m,m\rangle}$ over the distance $x=|n-m|$ for $W=1$, $u=0.5$ and $N=8500$. The two particle localization length $\xi_2(N)$ is extracted from the slope of the linear fit $y(x)=c-x/\xi_2(N)$.}
    \label{fig:A1}
\end{figure}

We explain in this appendix a step by step procedure for finding the localization length $\xi_2$ for a specific example with $W=1$, $N=8500$ and interaction strength $u=0.5$. We perform the two steps outlined in the main text in Sec.~\ref{sec:xi2u0}.  We compute
\begin{gather*}
    y(x) = ln\overline{\langle n,n|\tG|m,m\rangle}
\end{gather*}
averaged over $200$ disorder realisations. By definition (see Eq.~\eqref{eq:xi2-def}), $\xi_2(N)$ is extracted by fitting the  linear decay of $y(x)$ over the range of $x=|n-m|$, where $n=425(0.05N)$ and $m=8075(0.95N)$. This choice of the range was adapted from the Ref.~\onlinecite{frahm2016eigenfunction}: we discarded the parts close to the boundary to avoid its effect.  Figure~\ref{fig:A1} shows the average $y(x)$ and the linear fit $y(x) = c - x/\xi_2(N)$ for the chosen set of parameters. Similarly $\xi_2(N)$ are extracted from the inverse of the slope of the fits obtained for other $N\ge N_*(W=1)\approx 1500$.

\begin{figure}[h]
    \centering
    \includegraphics[width=0.49\textwidth,angle=0]{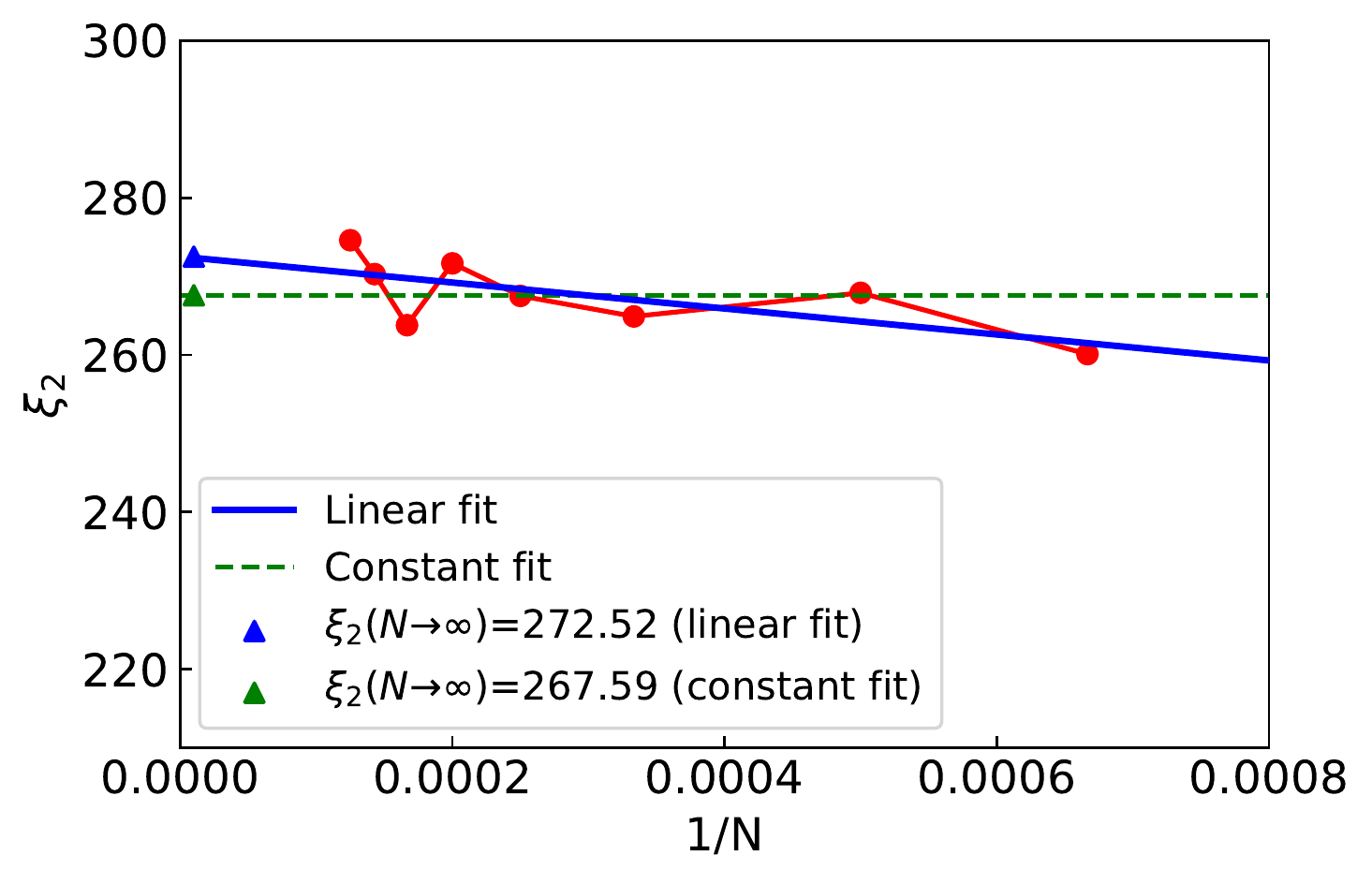}
    \caption{(Color online) Linear (solid blue) and constant (dashed green) fits to $\xi_2(N)$ for $W=1$ and $u=0.5$  to extract $\xi_2(N\to\infty)$. The blue and red trianlges indicate the extrapolated value for linear and constant fits respectively.}
    \label{fig:A2}
\end{figure}

Next we perform the extrapolation $N\to\infty$ of $\xi_2(N)$ to obtain $\xi_2(N\to \infty)$. We observe that $\xi_2(N)$ show only a mild dependence on $N$ for $N\ge N_*(W)$ and rather fluctuates around some average value. Therefore we try both a constant fit $f(x) = c$ and a linear fit $y = m*1/N + c$ from where $\xi_2(N\to\infty) = c$. The data and the fits are shown in Fig.~\ref{fig:A2}. The values of $\xi_2(N\to\infty)$ obtained from both fits are reasonably close, always within $5\%$. In the current setting example, $\xi_2 = 272.52$ (linear fit), $267.59$ (constant fit) thereby giving $\xi_2=267.59\pm4.9$. We used the constant fit systematically.

\bibliography{mbl}

\end{document}